# A concentric plasmonic platform for the efficient excitation of surface plasmon-polaritons

*Nancy Rahbany[1], Wei Geng[1], Rafael Salas-Montiel[1], Sergio de la Cruz[2], Eugenio R. Méndez[2], Sylvain Blaize[1], Renaud Bachelot[1], Christophe Couteau[1,3,4*]*

[1] Laboratory of Nanotechnology, Instrumentation and Optics, ICD CNRS UMR 6281, University of Technology of Troyes, 10000, Troyes, France

[2] División de Física Applicada, Centro de Investigación Científica y de Educación Superior de Ensenada, Carretera Ensenada-Tijuana No. 3918, Ensenada 22860, BC, México

[3] CINTRA CNRS-Thales-NTU, UMI 3288, Research Techno Plaza, 50 Nanyang Drive, Singapore

[4] Centre for Disruptive Photonics Technologies (CDPT), Nanyang Technological University, Singapore

*Correspondence and requests for materials should be addressed to C.C. (email: christophe.couteau@utt.fr).

**Abstract:** We propose a plasmonic device consisting of a concentric ring grating acting as an efficient tool for directional launching and detection of surface plasmon-polaritons (SPPs). Numerical simulations and optical characterizations are used to study the fabricated structured gold surface. We demonstrate that this circularly symmetrical plasmonic device provides an efficient interface between free space radiation and SPPs. This structure offers an excellent platform for the study of hybrid plasmonics in general and of plasmon-emitter couplings in particular, such as those occurring when exciting dye molecules placed inside the ring. As illustrated in this work, an interesting property of the device is that the position of excitation determines the direction of propagation of the SPPs, providing a flexible mean of studying their interactions with molecules or dipole-like emitters placed on the surface.

There is a growing interest nowadays in the study of light-matter interaction at the nanometer scale [1]. Using plasmonics, nanooptics and metamaterials, there is an important drive towards enhancing such an interaction. For instance, upon integrating single-photon sources with plasmonic structures, the emitter's enhanced Purcell factor in the weak coupling regime can be obtained [2, 3]. In plasmonics, researchers have proposed methods and devices that allow the understanding of SPP launching and propagation such as nanoslit arrays [4], metal and dielectric strips and wires [5, 6], carbon nanotubes [7], metallic nanowires [8, 9], metallic nanoparticles [10] and dislocated double-layer metal gratings [11]. Recent work has been done on combining quantum optics and plasmonics by studying the transmission of entangled photons between two metallic gratings upon exciting single waveguided SPPs [12]. The same platform can be used as an efficient nanoscale focusing device [13–15] as well as a confined surface plasmon polariton amplifier [16]. Plasmonic lenses made of concentric annular rings, known as bull's eye nanoantennas, have shown to increase the electric field enhancement and confinement [17, 18] and to control the emission direction of molecules placed in the center [19]. Furthermore, studies on integrating optical antennas into concentric ring gratings show that the field enhancement and coupling efficiency can be improved, as well as the fluorescent enhancement of emitters placed inside the ring-antenna system [20, 21]. Moreover, in order to get efficiently excited, dipole emitters must have a specific orientation with respect to the incident light polarization giving rise to a specific emission radiation pattern [22, 23]. There is thus an advantage in using a configuration that, unlike bull's eye nanoantennas, studies the behavior of propagating SPPs which can be used to excite emitters regardless of the emitters' orientation. Reciprocally, it would be useful if this structure would allow excited emitters placed inside it to generate SPPs that propagate and then reradiate in the far field out-coupling regardless of where they are located [21,22]. This is precisely what our present work intends to do by studying the excitation of SPPs with a specially fabricated ring grating structure. We demonstrate the interaction of these surface waves with fluorescent emitters placed inside the ring which makes our device useful in measuring the SPP emission direction. We also show that the generated SPPs propagate with a specific wave vector in a specific direction and exciting the ring grating from different positions leads to SPP orientation in a precise desired direction.

The article is organized as follows: The second section describes the design of the structures to couple a propagating beam of light into SPPs ("Design of the Structures" section). In "Experimental Results and Discussion" section, we present details of the fabrication and experimental results on the excitation and propagation of SPPs with the proposed structures. We also show the excitation of dye molecules by SPPs and the excitation of SPPs by the fluorescence of dye molecules using our plasmonic device. Finally, the last section contains our main conclusion ("Conclusion" section). We demonstrate that the proposed device can be used to study and control plasmon-emitter interactions, making the structure a convenient plasmonic platform.

**Design of the Structures**

In this section we present the procedure employed to design the structures and provide a brief description of the numerical approach used in this process, and in the evaluation of their performance.

Our structure is composed of a circular ring grating that provides a useful platform for focusing SPPs in its center where the wave amplitude reaches its maximum [14]. This allows studying SPP properties and their interactions with emitters and absorbers. A schematic illustration of the situation considered is illustrated in Fig. 1. If the radius of curvature of the rings is much greater than the diameter of the illuminating beam, the situation can be well approximated by the geometry of a linear grating assumed in the following calculations.

We start by considering the generic geometry illustrated in Fig. 2. A profile $\Gamma$ defines the interface between a metal with dielectric constant $\varepsilon_m(\omega)$ and air, with index of refraction $n_0=1$. We assume that the surface is invariant along $z$ and that the illumination is provided by a monochromatic beam of light propagating on the xy-plane.

Under the assumed conditions, for purely s- or p-polarized incident waves, the state of polarization is retained after interaction with the surface, and the two polarization modes may be treated independently. Since our primary interest is the excitation of SPPs, we only consider the case of p-polarization. Since in this case the magnetic field vector is along the $z$-direction, the problem is most naturally solved in terms of this component, which we represent by $H(x,y)$. We now provide a brief description of the numerical method employed to solve the problem. More details can be found in Refs. [24, 25].

We use the Green's integral theorem and consider the total field above the surface as being the sum of the incident and scattered fields.

By solving a set of integral equations [24, 25], we can then deduce the surface field. Once the surface field is found, it is straightforward to calculate the excitation efficiency. For this, let us consider the field associated with an SPP originating at $x = 0$, with amplitude $H_0$, traveling on a flat surface in the direction $+x$. This field can be expressed in the form

$$H^{(0)}(\mathbf{r}) = H_0 e^{ik_{SPP}x - \beta_0 y} \qquad y > 0, \qquad (1)$$

$$H^{(m)}(\mathbf{r}) = H_0 e^{ik_{SPP}x + \beta_m y} \qquad y < 0, \qquad (2)$$

for $x > 0$. Here,

$$k_{SPP} = k_0 \sqrt{\frac{\varepsilon_0 \varepsilon_m}{\varepsilon_0 + \varepsilon_m}} \qquad (3)$$

is the wavenumber of the SPPs, and

$$\beta_j = k_0 \sqrt{\frac{-\varepsilon_j^2}{\varepsilon_0 + \varepsilon_m}} \qquad (4)$$

is the decay constant of the field in the two media. In this last expression, the sub-index $j = 0$ for free-space or $j = m$ for metal, and $k_0 = \omega/c$ is the free-space wavenumber.

Evaluating the component of the Poynting vector that crosses the plane $x = 0$, and integrating along $y$, we find that [26]

$$P_x^{(j)} = L_z \frac{c^2}{8\pi\omega} \frac{H_0^2}{2\beta'_j} \Re e \left\{ \frac{k_{SPP}}{\varepsilon_j} \right\}, \qquad (5)$$

where $\beta_j' = \operatorname{Re}\{\beta_j\}$, $L_z$ is a length along the $z$ direction and, again, $j$ refers to air or metal.

Considering now a situation in which the SPP has been excited by the interaction between an illuminating beam and a surface feature, one can define the excitation efficiency as

$$\eta = \frac{P_x^{(0)} + P_x^{(m)}}{P_{inc}} \qquad (6)$$

where $P_{inc}$ is the power of the incident beam.

To design a coupling structure, we begin by considering the excitation efficiency that can be achieved with a sub-wavelength rectangular groove ruled on a metal surface. This efficiency is closely related to its electromagnetic resonances [27]. Once a suitable width and depth have been chosen, we consider a sequence of $N$ such groves placed in a regular grid, choosing the period $d$ in such a way that the SPPs excited by them interfere constructively. This means that the periodicity of the grating must be such that the following matching equation is satisfied

$$k_{SPP} = n_0 k_0 \sin\theta \pm q k_g \qquad (7)$$

where $k_g = 2\pi/d$ is the grating wavenumber, and $q$ is an integer.

To maximize the efficiency, and reduce the effects of stray light, it is convenient to choose a period that is smaller than the wavelength. In such circumstances, apart from the specular order, the grating produces no propagating orders for small angles of incidence. For a wavelength of 980 nm and a period of 800 nm, for example, Eq. (7) predicts that the coupling condition is reached when θ is about -10°.

Long gratings, however, do not couple efficiently to freely propagating SPPs [26, 27]. This is because, as the coupled SPPs propagate through the grating, they are diffracted by the grating and radiate into air and the metal. Through numerical studies with a variable number of grooves, we have found that there is practically no improvement of the efficiency after five grooves [27]. The designed couplers consist then of five grooves.

The inset of Fig. 3 shows a schematic diagram of the grating coupler. The position of a Gaussian illuminating beam is also indicated in that figure. To excite SPPs to the right of the structure the beam must come downward from right to left (negative angle of incidence). An efficiency map, as a function of the position of the beam with respect to the center of the structure and the angle of incidence, is shown in Fig. 3. The calculations were carried out using a home-made code based on the integral equation approach explained briefly above. The wavelength used is $\lambda$ = 980 nm and the experimentally determined dielectric constant for gold was $\varepsilon_m$ = -41.2+1.54 $i$. The 1/$e$ value of the radius of the focused Gaussian beam was $g$ =196 nm. The period of the five groove grating is $d$ = 820 nm, the depth of the grooves $h$ = 196 nm, and their widths $w$ = 416.5 nm.

One can see that this compact structure can couple, under the right conditions, about 40% of the incident power into freely propagating SPPs. It is also observed that the coupling is fairly tolerant to changes in the angle of incidence and the position of the beam. For instance, a change of as much as 5° in either direction from the optimum angle of incidence, reduces the coupling efficiency by about 5%. Changing the position of the incident beam by about 0.5 μm from the optimal $x_c$ = 0.7 μm has a similar effect.

Figure 4a,b presents numerical simulations of SPPs propagation within our platform through calculated maps of $|H(x,y)|^2$ in the near field of the surface structure, under different conditions of excitation. In Fig. 4a, we illustrate the excitation of SPPs by an external beam and the subsequent conversion of the excited SPPs into propagating waves. The illumination is provided by a Gaussian beam that makes an angle of incidence θ = -10°. The rest of the parameters are as in the efficiency map. We note in the figure the Wiener-like interference fringes produced by the interference between the incident and specularly reflected beams, as well as the evanescent nature of the excited SPPs and their interaction with the second grating. This figure illustrates the fact that these short gratings are not only efficient couplers, but are also useful to couple out SPPs for their far-field detection. In Fig.4b, we illustrate a case in which the SPPs are excited by a vertical point dipole on the

surface. The excited SPPs travel along the surface and radiate upon interaction with the lateral gratings and couple out to the far field.

**Experimental Results and Discussion**

The structures fabricated for this work consisted of five concentric grooves that, when illuminated by a laser beam, result in the excitation of a SPP that propagates radially across the circular structure and reaching the opposite side of the ring (see Fig. 1). Ring gratings with periods of 520 nm and 800 nm, that are compatible with the 632.8 nm and 980 nm laser sources used, and with diameters varying from 50 to 400 μm, were fabricated for this work.

To fabricate the samples, Si substrates were spin-coated with a 200 nm layer of PMMA (4000 rpm). The grooves were etched using Electron Beam Lithography (EBL) techniques. The EBL process was followed by the evaporation of a 3 nm layer of chromium and a 150 nm layer of gold. As this last layer is optically thick, the sample can be assumed to consist of gold only. The circular grating whose image is shown in Fig. 5 was fabricated with a dose of 90 μC/cm$^2$. This particular diameter of the circle is about 302 μm and the period is 816 nm, which is compatible for exciting with a laser of 980 nm wavelength (as shown in Fig. 1). The slight difference in the size of the grating grooves and the size of the engraved slits is due to the dose of the electron beam applied (only few percent difference).

The structures were studied using an optical set-up in which an optical fiber was used to illuminate the ring grating from the top (see Fig. 1) and excite SPPs that propagate along the ring diameter and decouple out at the opposite end of the circle. In Fig. 6, Lycopodium powder of micron-size dust-like type was spread on the surface of a ring grating of 800 nm period and 100 μm diameter. An incident laser of 980 nm wavelength is used to excite the ring at a certain position. The propagating SPPs are scattered by the Lycopodium powder and are shown to propagate radially along the ring diameter until they couple out again at the opposite position on the grating. The powder is used to scatter SPPs to photons in the far-field, thus allowing us to observe their directional propagation.

The propagation length $L$ of SPPs on a flat surface ($L = 1/2k_{SPP}^{"}$ where $k_{SPP}$ is the complex SPP wavenumber given in Eq. (7)) was studied using fabricated samples with circles that have diameters ranging from 200 μm to 400 μm in 50 μm increments. For this, a small section of the circular grating was illuminated to excite SPPs, and we measured the radiated optical power at the other end of the circle. The results are shown in Fig. 7. One observes that the power decreases exponentially as a function of the diameter. An exponential decay function $I = I_0 e^{-x/L}$ was used to fit the experimental data and obtain the propagation length of the SPPs. We found that $L_{exp}$ = 69.8 μm. This value is significantly different than the one obtained by simulations using the experimentally measured index ($L_{sim}$ = 166 μm) mostly due to surface roughness that scatters the SPPs at the surface (see Fig. 6) and increases absorption.

The fabricated ring structures can also be used to study the interaction of SPPs with surface structures, such as rectangular grooves ruled across the diameter of the ring. As illustrated in Fig. 8 (a), such grooves constitute, effectively, SPPs beamsplitters. In order to identify the position of the incident laser spot as well as that of the output reflected and transmitted spots, we engraved by e-beam lithography additional marks close to the circumference of the ring corresponding to the polar angles every 10° in the plane of the ring grating, as shown in Fig. 8 (b). We then recorded the intensity of the transmitted and reflected spots as a function of the incident laser spot position. This is achieved by moving the incident laser spot around the circle to excite SPPs at different positions. The results are shown in Fig. 8 (c), where one observes the expected complementarity between the reflected and transmitted intensities. Therefore, understanding the propagation properties of surface plasmons in our structure allows us to extend this study to observe plasmon scattering by surface defects that require non-perpendicular excitation. Controlling the direction of propagation of surface plasmons helps us address specific defects present on the surface and acquire valuable information related to their optical properties. This can be useful for important applications in surface-roughness analysis, biosensing, and photonic nano-device development.

In addition, we studied the effect of the polarization on the coupling efficiency by varying the polarization of the incident light from perpendicular to parallel with respect to the grating grooves. The measured output power as a function of the polarization angle is shown in Fig. 9. We observe that the maximum coupling efficiency occurs when the incident light is p-polarized, i.e. the electric field is orthogonal to the grating grooves, which corresponds to TM polarization. The curve displays a $\cos^2$ behavior that reaches its minimum value for s-polarized light (TE polarization parallel to the grating grooves). FDTD simulations carried out with Lumerical software are also shown in Fig. 9, confirming the experimental observation. We point out, however, that for the simulation we added a constant background level observed experimentally even for the TE polarization case. This background is most likely caused by scattered light and is responsible for the low contrast observed experimentally.

Finally, in order to illustrate the potential of our plasmonic platform for hybrid plasmonics, we studied the plasmon-emitter coupling by propagating SPPs that excite emitters placed in the center of the ring. We choose to work with Atto 633 dye molecules purchased from the company Atto-Tec which we characterize by measuring the absorption and photoluminescence (PL) spectra as shown in Fig. 10. The absorption of a 6.66 mg/L solution of Atto 633 dyes is measured using a Cary 100 UV-Visible absorption spectrophotometer from Varian. The maximum absorption wavelength obtained is at 630 nm. The photoluminescence of such dyes on a glass substrate is measured by using a home-built confocal microscopy system. The maximum emission peak was observed at 655 nm. Consequently, in order to achieve SPP-emitter coupling via our structure, a 1 µm drop of Atto 633 dye molecules of 0.33 g/L concentration was placed inside a 50 µm ring grating (without beamsplitter) with a period of 525 nm with the help of micromanipulators from Imina Technologies with micrometer size tungsten probe tips. Illuminating the ring grating with a 632.8 nm laser produces SPPs traveling along the ring diameter. As the Atto dyes have a matching absorption wavelength, these SPPs can efficiently excite them.

As observed in the inset of Fig. 11, Atto dyes form clusters that are a bit displaced from the exact center of the ring. Sample observation as well as photoluminescence spectroscopy is performed using a home-built confocal microscope system of high sensitivity, including Peltier-cooled CCD camera (T = -80°C) and a 50X, NA= 0.95 objective. A continuous-wave laser spot ($\lambda$ = 632.8 nm) is fixed at the ring grating circumference and the PL spectrum is measured exactly and exclusively at the position of the dyes by spatially selecting the detection area desired using a pinhole placed in front of the spectrometer. We present in Fig. 11 the PL spectrum of the Atto dyes which shows a peak at 656 nm confirming that the dyes are successfully excited. The output spot observed on the opposite side of the ring is due to the fact that not all of the incident light is absorbed by the dyes, which have a quantum yield of 64% (Fig. 10). In addition, using a time-correlated single photon counting setup, the lifetime of the Atto dyes was measured on a glass substrate, on a gold substrate, and inside a 50 μm ring grating (lifetimes not shown). The obtained ratio of their lifetime on glass to that on gold was found to be 2.5. Similar studies reported in the literature on the lifetime of ATTO655-SA on glass and on a developed silver–gold nanocomposite (Ag–Au–NC) substrate have found a glass to Ag-Au-NC ratio of ~3.1 [28]. However, we observed no change in lifetime for dyes in the center of the ring grating compare to unstructured gold surface only. This is expected as the ring size is too big to start acting as a cavity for enhancing the emitters' fluorescence. Only the propagation behavior of SPPs is tackled with our device.

As already mentioned, the reciprocal case has also been studied (see Fig. 4b for numerical simulations of a point dipole coupled to SPPs and re-radiating into photons via the gratings). Here, an external laser source is used to excite the emitters directly in the center of the ring, which subsequently act as SPP generators themselves. These SPPs propagate from the center of the ring and couple out from the ring grating. Due to the random orientation of the dye molecules in the center and their random distance from the metal surface, SPPs propagate at arbitrary uncontrolled angles from the center and couple out upon reaching the ring grating. The emission of the dyes is composed of radiative emission in the far field, guided emission into SPP modes, observed through the radiation from the ring, and non-radiative emission. As seen in the inset of Fig. 12, some dye molecule clusters are not placed exactly in the center of the ring. Thus the circumference is not uniformly illuminated and several bright spots are observed. As seen in Fig. 9, SPPs can only be generated along p-polarization but there is also a wave vector filtering caused by the grating when SPPs are generated inside the ring. Indeed, Eq. (7) imposes that the generated SPPs from the dye clusters have a wave vector perpendicular to the grating grooves in order to couple out at the ring circumference. If the wave vector is not radial then no outcoupling is possible. From this feature, one can extrapolate the position of the clusters with respect to the ring center. This further reinforces the idea of using our device with single dipoles in the future. We demonstrated that the out-coupled light at the rings originates from the excited emitters by measuring the photoluminescence spectra recorded at several positions on the ring circumference. An example spectrum is shown in Fig. 12, which coincides with the fluorescent spectrum of the dye and clearly displays a peak at 657 nm. We note that in the inset of Fig. 12, the observed vertical white line is a result of the high intensity of the laser spot that saturates the imaging camera and could not be filtered out.

**Conclusion**

We have proposed a promising device made of a concentric plasmonic ring grating for launching, manipulating and visualizing surface plasmon polaritons. In particular, we have shown that it can be successfully used to study emitter-plasmon couplings. Experimental and numerical work was carried out to study the dependence of the excitation on the angle of incidence and the polarization of incident light. We also determined experimentally the SPP propagation length. In addition, SPPs coupled through the ring grating were used to excite emitters placed at the center of the ring. We also presented work where exciting the emitters directly leads to the launching of SPPs that couple to the grating and reemit light into the far field.

Further work can be accomplished with our rotationally symmetrical plasmonic structure, including placing single emitters in the ring and integrating them with optical nanoantennas. After being excited by a light source, single molecules can themselves act as SPP generators. The direction of propagation of the generated SPPs depends on the orientation of the single molecule with respect to the incident light polarization, as well as its position in height with respect to the metal surface. Exciting single emitters with p-polarization with a well-known orientation in the center will allow us to control the SPP propagation and orient it in a specific desired direction within the ring. In addition, the advantage of having a concentric structure is that it could be used to excite emitters from anywhere on its circumference depending on the emitters' spatial orientation and position, which would allow us to control and orient the SPP propagation in a desired direction. Furthermore, focusing and manipulating SPP propagation can be achieved by observing plasmon scattering by surface defects which is a very important aspect in studying the properties of surface plasmon polaritons. Hence, in addition to being a source for SPP launching, we show that our plasmonic device can be used to study and control plasmon-emitter coupling, making the structure an efficient plasmonic platform and thus opening up new possibilities for studying the interaction of SPPs with single dipoles on substrates.

**Acknowledgment:** The authors would like to thank A. Bruyant and G. Colas des Francs for helpful discussions, as well as S. Kostcheev for technical support for the fabrication. N.R. would like to thank the French Ministry of Education for her PhD grant. The authors thank the region Champagne-Ardenne platform Nanomat for fabrication and characterization facilities. C.C. and S.B. would like to acknowledge the financial support by the CNRS PEPS project 'InteQ'. C.C. thanks the partial funding by the Champagne-Ardenne region via the project 'NanoGain' and S.B., R. S.-M. and C.C. thank the partial support by the French ANR via the project 'SINPHONIE.' The work of S. de la C. and E.R.M. was supported in part by CONACYT, under grant 180654.

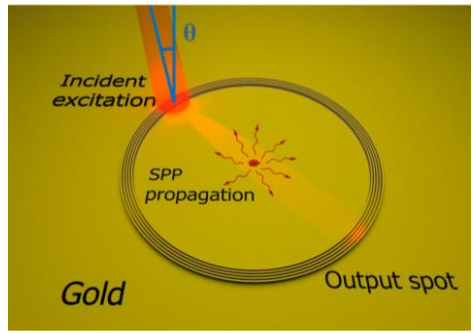

**Fig. 1** Sketch of the ring grating structure illuminated with a beam incident at angle θ. SPP propagation is shown along the ring diameter which excites emitters placed in the center and decouples upon reaching the grating.

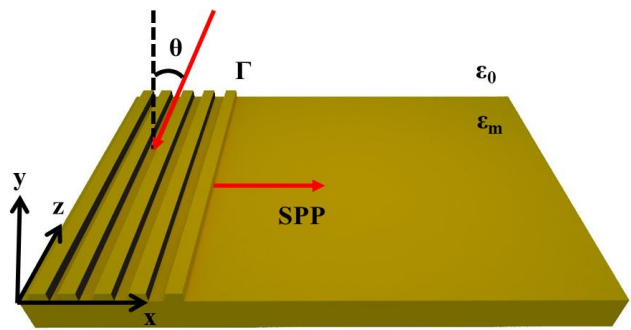

**Fig. 2** Schematic diagram of the geometry considered. A beam of light illuminates the corrugated section of a metallic surface, exciting a SPP that propagates along the flat section. The angle of incidence θ is measured from the *y*-axis in the counterclockwise direction; it thus has a negative value in the figure.

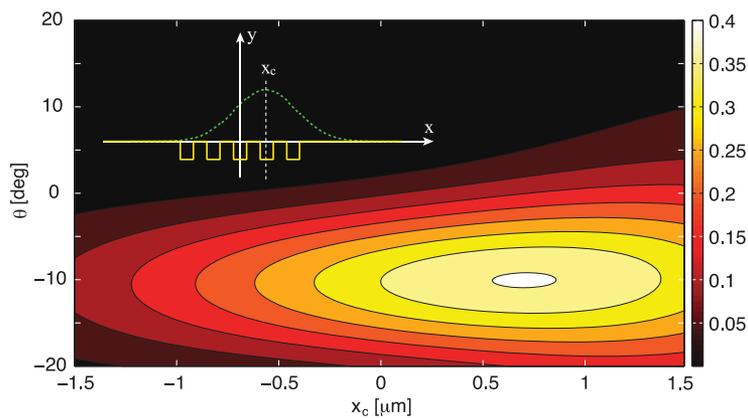

**Fig. 3** Efficiency of excitation of SPPs as a function of the angle of incidence θ and the position of the center of the beam on the surface $x_c$. The inset illustrates the surface profile structure and the position of the incident Gaussian beam on the surface.

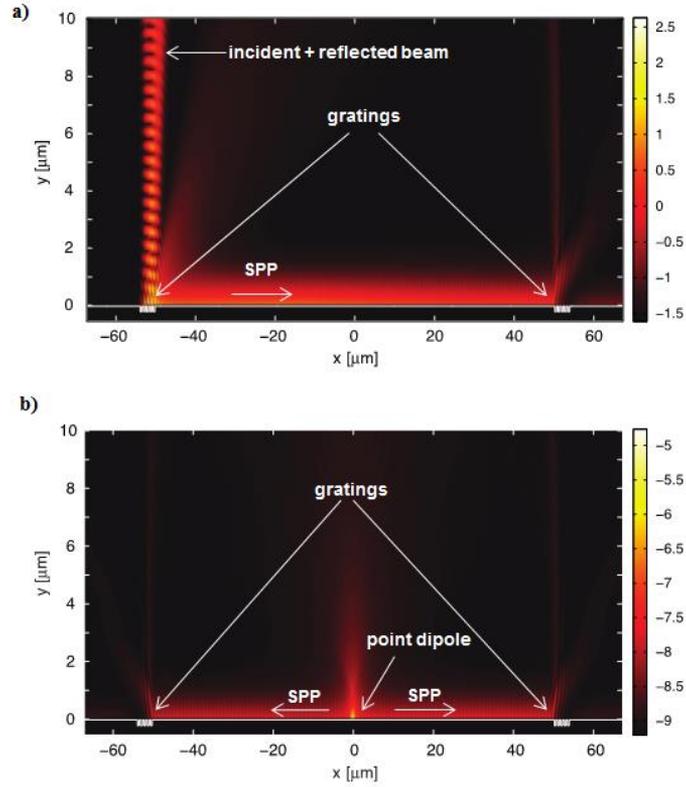

**Fig. 4** Calculated near field magnetic intensity ($|H(x,y)|^2$) map of: a) the excitation of SPPs through the interaction of a Gaussian beam with a grating coupler, and b) the excitation of SPPs by a point dipole on the surface and their subsequent conversion into volume waves upon interaction with the gratings.

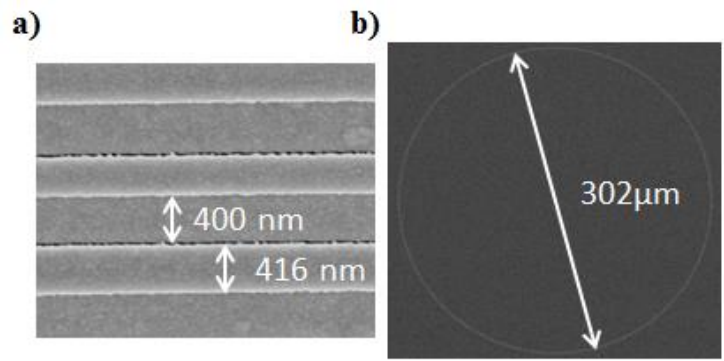

**Fig. 5** SEM images showing the (a) period and (b) diameter of a ring grating.

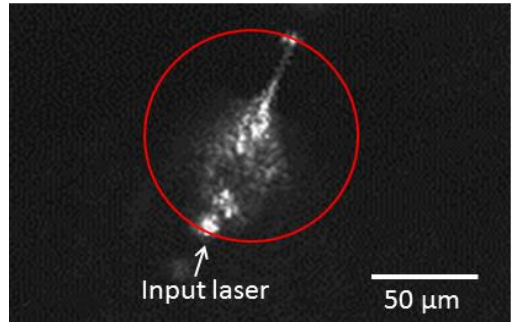

Fig. 6. Optical image showing the SPP propagation along the ring diameter scattered in the far-field by the lycopodium powder.

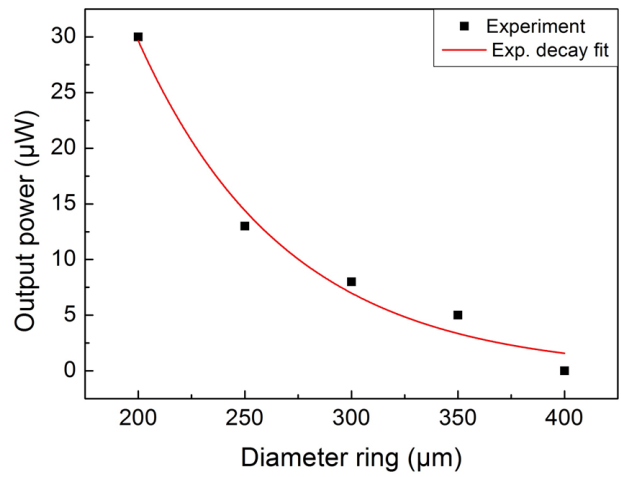

**Fig. 7** Experimentally measured output power as a function of the ring diameter.

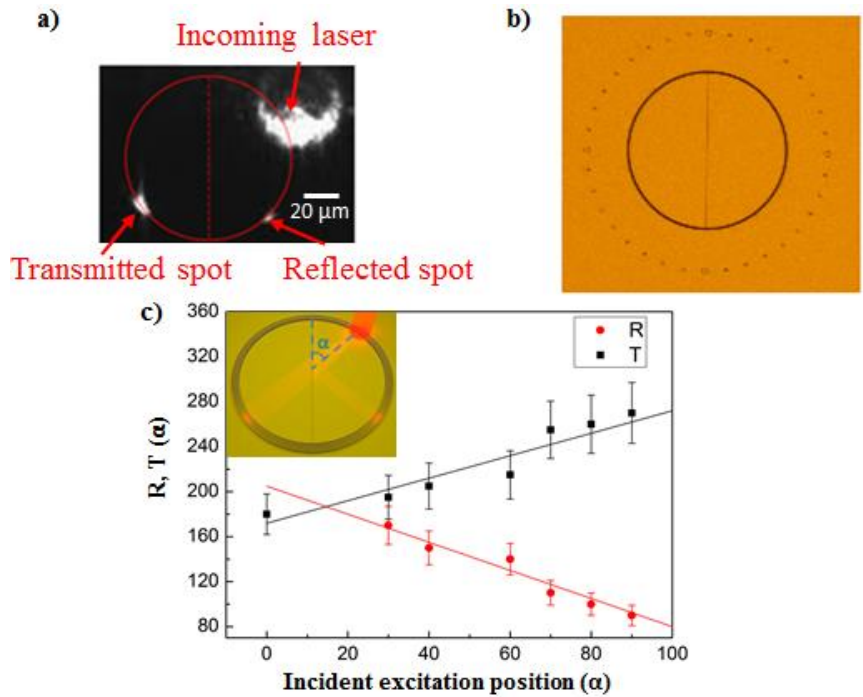

**Fig. 8** (a) SPPs propagation on a ring with a beamsplitter (b) Optical image showing marks done on a ring grating with a beam splitter to identify the position of the incident, reflected, and transmitted spots. (c) Intensity of the reflected and transmitted spot positions as a function of the incident excitation position for a ring with beamsplitter. The angle α between the BS and the measured spot position is chosen as the unit of both axes.

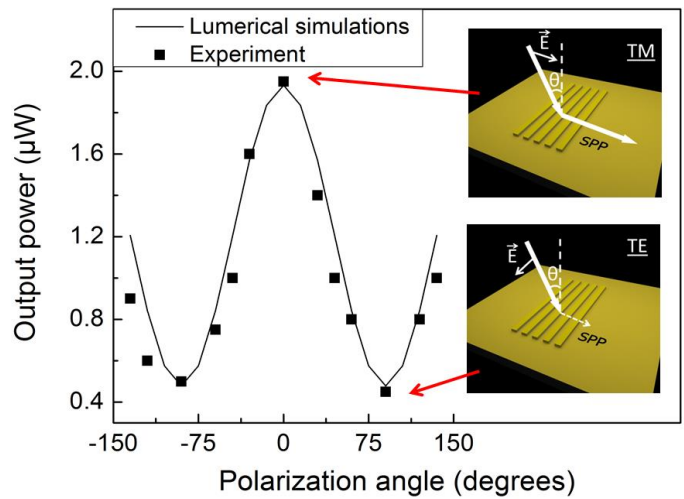

**Fig. 9** Output power as a function of the incident polarization: experiment and simulation

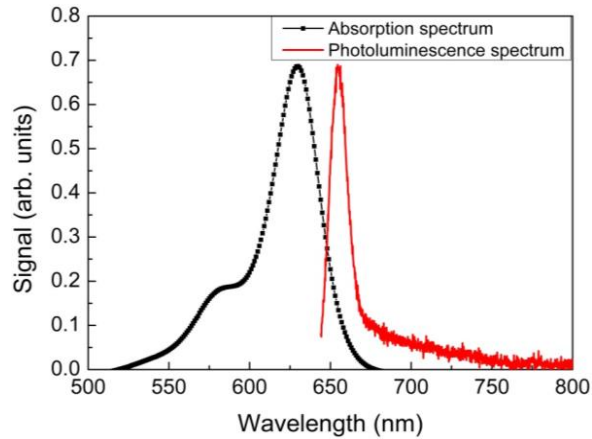

**Fig. 10** Absorption and photoluminescence spectrum of Atto 633 dyes.

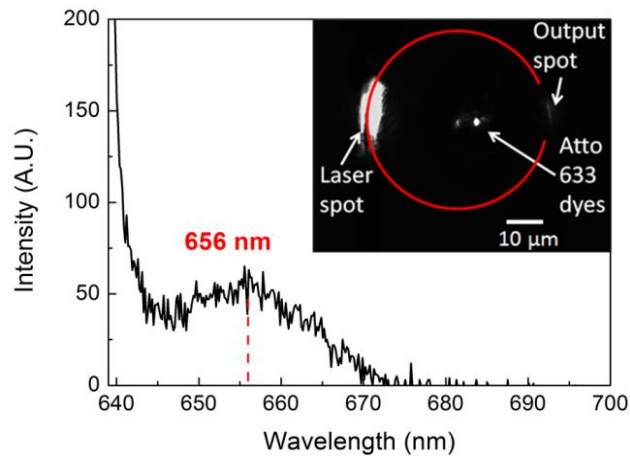

**Fig. 11** Atto 633 dyes excited by SPPs generated by the grating. Photoluminescence spectrum and top view of the emitted optical intensity.

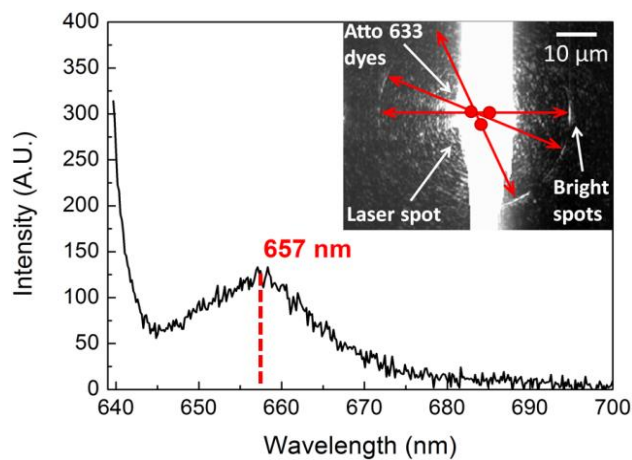

**Fig. 12** Photoluminescence spectrum of an output spot on the ring caused by SPPs generated by exciting Atto 633 dyes directly by the laser. Inset: top view of the involved emitted light.